\documentclass[twocolumn,showpacs,preprintnumbers,amsmath,amssymb,superscriptaddress]{revtex4}

\usepackage{graphicx}
\usepackage{graphicx}
\usepackage{here}
\begin{document}
\title{Temperature-dependent soft x-ray photoemission and absorption 
studies of charge disproportionation in La$_{1-x}$Sr$_x$FeO$_3$}

\author{H. Wadati}
\email{wadati@wyvern.phys.s.u-tokyo.ac.jp}
\affiliation{Department of Physics and Department of Complexity 
Science and Engineering, University of Tokyo, 
Kashiwanoha 5-1-5, Kashiwa, Chiba 277-8561, Japan}

\author{A. Chikamatsu}
\affiliation{Department of Applied Chemistry, University of Tokyo, 
Bunkyo-ku, Tokyo 113-8656, Japan}

\author{R. Hashimoto}
\affiliation{Department of Applied Chemistry, University of Tokyo, 
Bunkyo-ku, Tokyo 113-8656, Japan}

\author{M. Takizawa}
\affiliation{Department of Physics and Department of Complexity 
Science and Engineering, University of Tokyo, 
Kashiwanoha 5-1-5, Kashiwa, Chiba 277-8561, Japan}

\author{H. Kumigashira}
\affiliation{Department of Applied Chemistry, University of Tokyo, 
Bunkyo-ku, Tokyo 113-8656, Japan}

\author{A. Fujimori}
\affiliation{Department of Physics and Department of Complexity 
Science and Engineering, University of Tokyo, 
Kashiwanoha 5-1-5, Kashiwa, Chiba 277-8561, Japan}

\author{M. Oshima}
\affiliation{Department of Applied Chemistry, University of Tokyo, 
Bunkyo-ku, Tokyo 113-8656, Japan}

\author{M. Lippmaa}
\affiliation{Institute for Solid State Physics, University of Tokyo, 
Kashiwanoha 5-1-5, Kashiwa, Chiba 277-8581, Japan}

\author{M. Kawasaki}
\affiliation{Institute for Materials Research, Tohoku University, 
2-1-1 Katahira, Aoba, Sendai 980-8577, Japan}

\author{H. Koinuma}
\affiliation{Materials and Structures Laboratory, Tokyo Institute of
Technology, 4259 Nagatsuta, Midori-ku, Yokohama 226-8503, Japan}

\date{\today}
\begin{abstract}
We have measured the temperature dependence of the photoemission and 
x-ray absorption spectra of La$_{1-x}$Sr$_x$FeO$_3$ (LSFO) 
epitaxial thin films with $x=0.67$, where charge disproportionation 
($3\mbox{Fe}^{3.67+}\rightarrow 2\mbox{Fe}^{3+}+
\mbox{Fe}^{5+}$) resulting in long-range spin and charge ordering 
is known to occur below $T_{CD}=190$ K. 
With decreasing temperature we observed gradual changes of the spectra 
with spectral weight transfer over a wide energy range of $\sim 5$ eV. 
Above $T_{CD}$ the intensity at the Fermi level ($E_F$) 
was relatively high compared to that below $T_{CD}$ 
but still much lower than that in conventional metals. 
We also found a similar temperature dependence for $x=0.4$, and to a 
lesser extent for $x=0.2$. 
These observations suggest that a local charge disproportionation occurs 
not only in the $x=0.67$ sample below $T_{CD}$ but also over a wider 
temperature and composition range in LSFO. This implies that 
the tendency toward charge disproportionation 
may be the origin of the unusually wide insulating region of the 
LSFO phase diagram. 
\end{abstract}
\maketitle
\section{Introduction}
It has been well known that carrier doping into Mott insulators 
causes various intriguing physical 
phenomena \cite{rev}. Among them, charge ordering (CO) and associated metal-insulator 
transition (MIT) have attracted particular interest in relation to charge stripes in
high-$T_c$ cuprates \cite{stripe} and spin-charge-orbital ordering in 'colossal' 
magnetoresistive manganites. 
La$_{1-x}$Sr$_x$FeO$_3$ (LSFO) with $x\simeq 2/3$ exhibits a charge
disproportionation (CD) 
from the average valence state of $\mbox{Fe}^{3.67+}(d^{4.33})$ into 
$2\mbox{Fe}^{3+}(d^5)+\mbox{Fe}^{5+}(d^3)$, 
a unique type of CO, below $T_{CD}=190$ K \cite{Takano}, 
accompanied by both a resistivity jump by an order of magnitude 
and antiferromagnetic ordering. 
A neutron diffraction study by Battle {\it et al.} \cite{Battle} revealed a 
spin-density wave (SDW) of sixfold periodicity and a charge-density 
wave (CDW) of threefold periodicity along the $\langle 111 \rangle$
direction. They also 
reported the apparent absence of structural distortion in the charge 
ordered state, whereas 
electron diffraction study by Li {\it et al.} \cite{Li} revealed 
superspots corresponding to a local structural distortion along the $\langle 111 \rangle$ 
direction for the sample of $x=0.7$. 
Optical conductivity spectra showed a splitting of  
optical phonon modes due to the folding of the phonon dispersion 
branch, caused by the lattice distortion below $T_{CD}$ \cite{Ishikawa}. 
Cluster-model analyses of photoemission spectra revealed that 
the charge disproportionation 
is better described as 
$3d^5\underline{L}^{0.67} \rightarrow 2d^5+d^5\underline{L}^2$, 
where $\underline{L}$ denotes a hole in the O $2p$ band, 
rather than $3d^{4.33} \rightarrow 2d^5+d^3$, because of the 
oxygen $p$ character of holes doped into the charge-transfer 
insulator LaFeO$_3$ \cite{Bocquet}. 
Matsuno {\it et al.} \cite{Matsuno} studied temperature 
dependent changes near the Fermi level ($E_F$) in the photoemission 
spectra of LSFO with $x=0.67$ and found that 
the intensity at $E_F$ clearly changed across the transition temperature. 
They also reported smaller but finite changes for 
$x=0.55$ and 0.80 and suggested that a local charge
disproportionation may occur even away from $x=0.67$. 
It was recently reported that M\"{o}ssbauer spectra of 
LSFO with $x = 0.5$ \cite{Yoon} 
show temperature-dependent valence changes 
similar to $x = 0.67$.

In the previous work, 
we have performed soft x-ray photoemission and absorption 
measurements of high-quality La$_{1-x}$Sr$_x$FeO$_3$ single crystals 
epitaxially grown by the pulsed laser deposition (PLD) method 
{\it in situ} \cite{Wadati}. 
Owing to the atomically flat, well-defined 
surfaces of the single-crystal thin films and the use of high energy 
soft x-rays ($h\nu =600-710$ eV) compared to the previous work 
($h\nu =21.2-100$ eV) \cite{Matsuno}, 
the photoemission spectra, especially of the valence band, 
revealed well-resolved structures directly related 
to the Fe $3d$ states, 
enabling detailed electronic structure 
studies of this material. 
In the present work, we address the questions of how the electronic
structure of LSFO changes when the CD occurs as a function of temperature, 
using soft x-ray photoemission and absorption. 
We have also utilized Fe $2p\rightarrow 3d$ resonant 
photoemission effect to further enhance 
the Fe $3d$ contribution. 
We have investigated LSFO with $x=0.67$ to observe the
change of the electronic structure accompanied by charge 
disproportionation, as well as $x=$ 0, 0.2, and 0.4 samples.
\section{Experiment}
\begin{figure}
\begin{center}
\includegraphics[width=7cm]{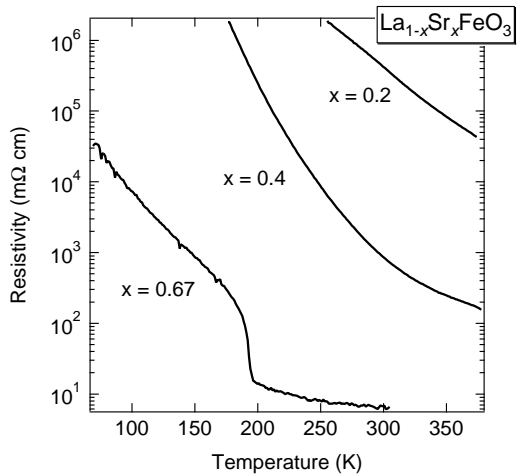}
\caption{Electrical resistivity of La$_{1-x}$Sr$_x$FeO$_3$ 
thin films.}
\label{pt}
\end{center}
\end{figure}

The experiment was performed at BL-2C and BL-1C of Photon Factory (PF), 
High Energy Accelerators 
Research Organization (KEK), using a combined 
laser molecular beam epitaxy (MBE) photoemission 
spectrometer system. 
The experimental setup is the same as that in Ref.~\cite{Wadati} 
and described in detail in Ref.~\cite{Horiba}. 
Epitaxial thin films of LSFO were fabricated 
from ceramic targets of desired chemical compositions 
by the PLD method. 
Single crystals of Nb-doped SrTiO$_3$ were used as substrates. 
Nb doping was necessary to avoid charging effects during the PES 
measurements. 
A Nd:YAG laser was used for ablation in its frequency-tripled mode 
($\lambda=355$ nm) at a repetition rate of 0.33 Hz. 
The substrates were annealed at 1050${}^{\circ}$C at an oxygen pressure of 
$\sim 1\times 10^{-6}$ Torr to obtain an atomically flat TiO$_2$-terminated 
surface \cite{kawasaki}. 
LSFO thin films of $\sim$ 100 monolayers were deposited on the
substrates at 
950${}^{\circ}$C at an oxygen pressure of 
$\sim 1\times 10^{-4}$ Torr. The films were post-annealed at 400${}^{\circ}$C 
at an atmospheric pressure of oxygen to remove oxygen vacancies.
The samples were then transferred 
from the MBE chamber to the spectrometer 
under an ultrahigh vacuum. 
The surface morphology of the measured films was checked 
by {\it ex-situ} atomic force microscopy. 
The electrical resistivity of $x=0.67$ sample showed 
a jump at 190 K ($=T_{CD}$), 
whereas the other samples showed insulating behaviors 
in the whole temperature range, as shown in Fig.~\ref{pt}.

Photoemission measurements were performed under an ultrahigh vacuum 
of $\sim 10^{-10}$ Torr from room temperature down to 10 K. 
The PES spectra were measured using 
a Scienta SES-100 electron-energy analyzer. 
The total energy resolution was about 200 meV at the photon energy of 
600 eV, about 250 meV at that of 710 eV 
(on Fe $2p\rightarrow 3d$ resonance), and about 35 meV 
at that of 40 eV. 
The $E_F$ position was determined by measuring gold 
spectra. 
The XAS spectra were measured in the total-electron-yield mode. 
The stoichiometry of the thin films was characterized 
by analyzing the relative intensities of relevant core levels, 
confirming that the compositions of the samples were nearly those of the 
ceramic targets \cite{Wadati}.
\section{Results and discussion}
\subsection{La$_{0.33}$Sr$_{0.67}$FeO$_3$}
\begin{figure}
\begin{center}
\includegraphics[width=8cm]{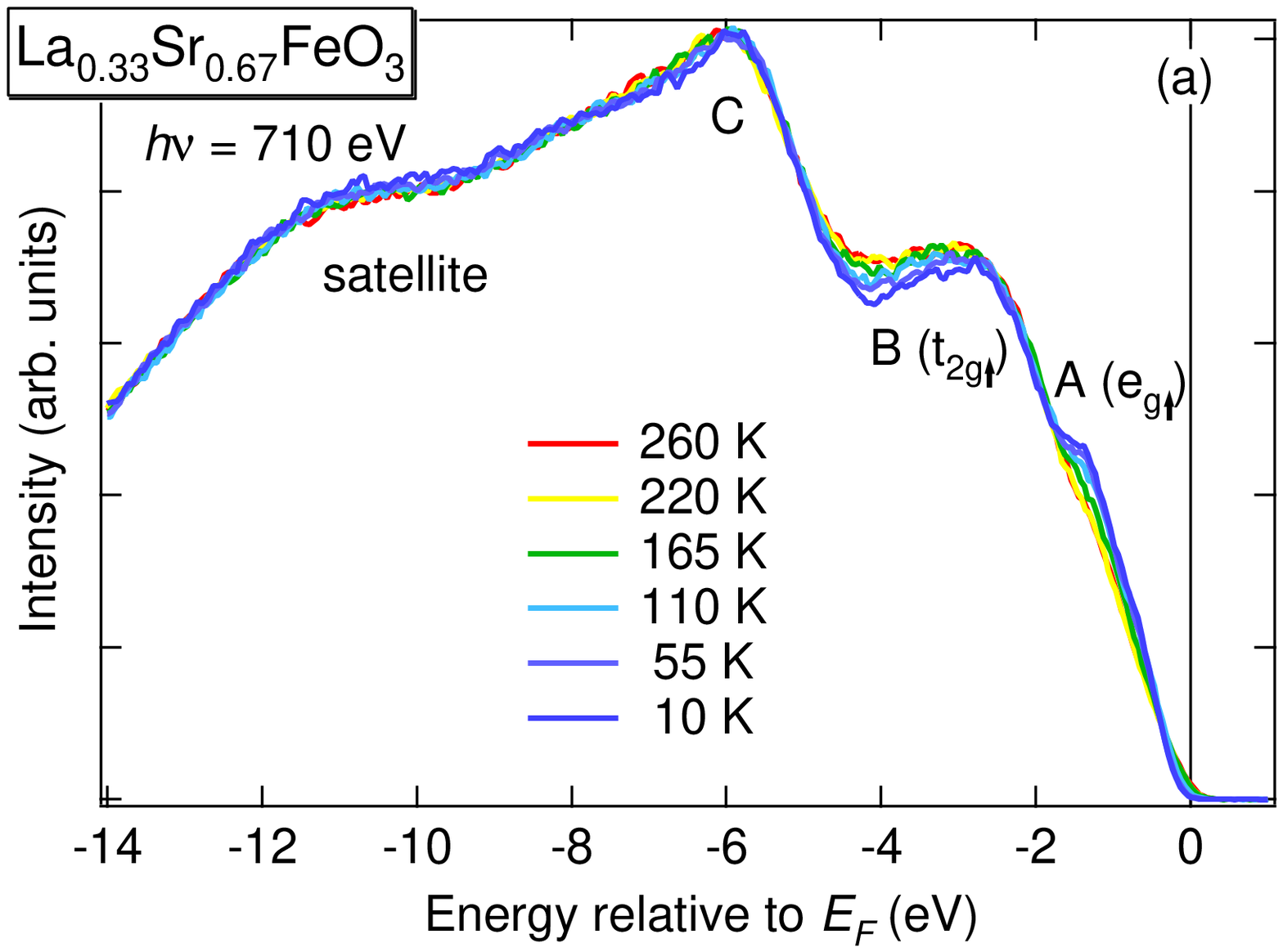}
\includegraphics[width=8cm]{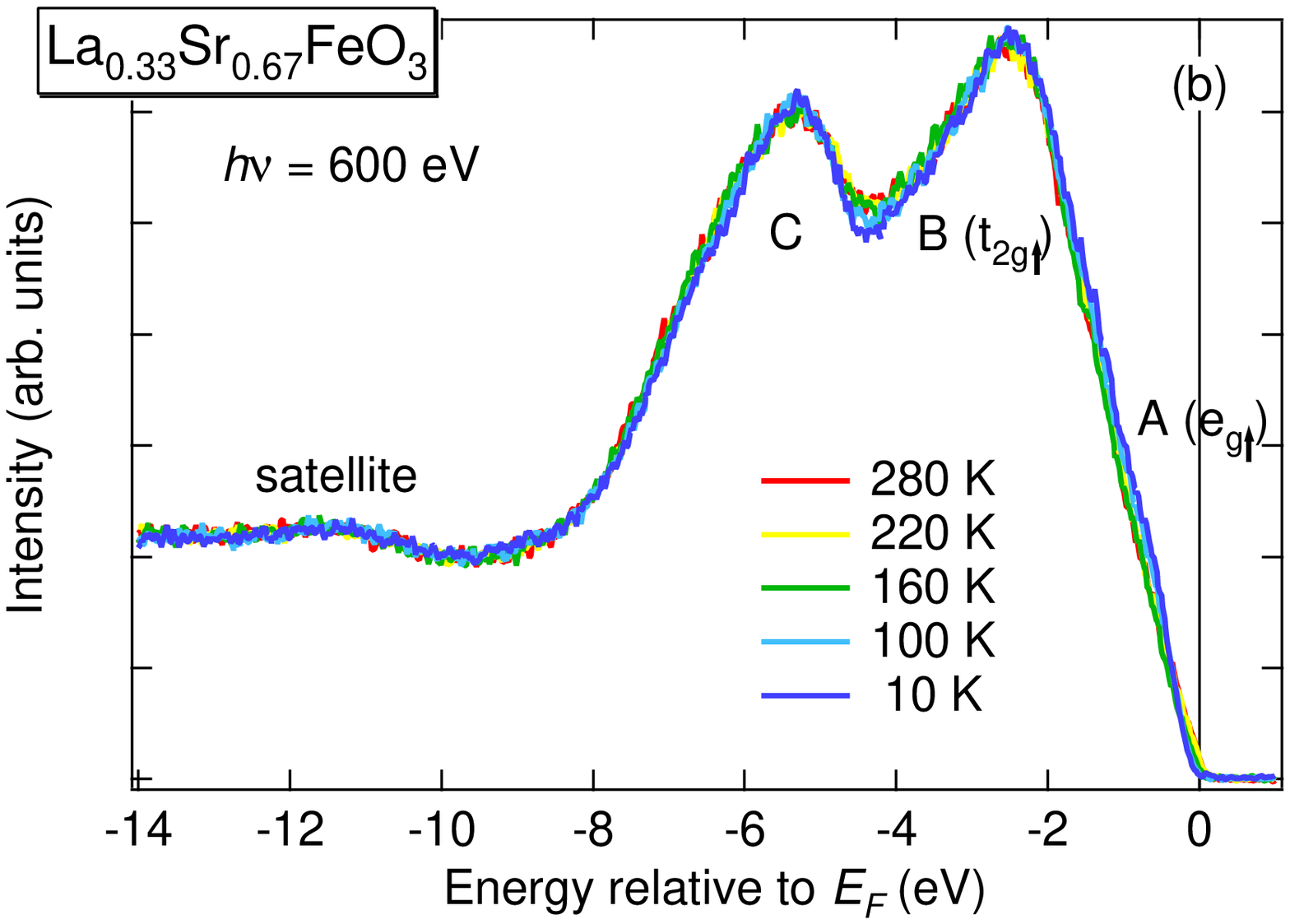}
\caption{(Color) Temperature dependence of the valence-band photoemission 
spectra of La$_{0.33}$Sr$_{0.67}$FeO$_3$. 
(a): $h\nu = 710$ eV (on Fe $2p\rightarrow 3d$ resonance), 
(b): $h\nu = 600$ eV.} 
\label{0.671}
\end{center}
\end{figure}
\begin{figure}
\begin{center}
\includegraphics[width=9cm]{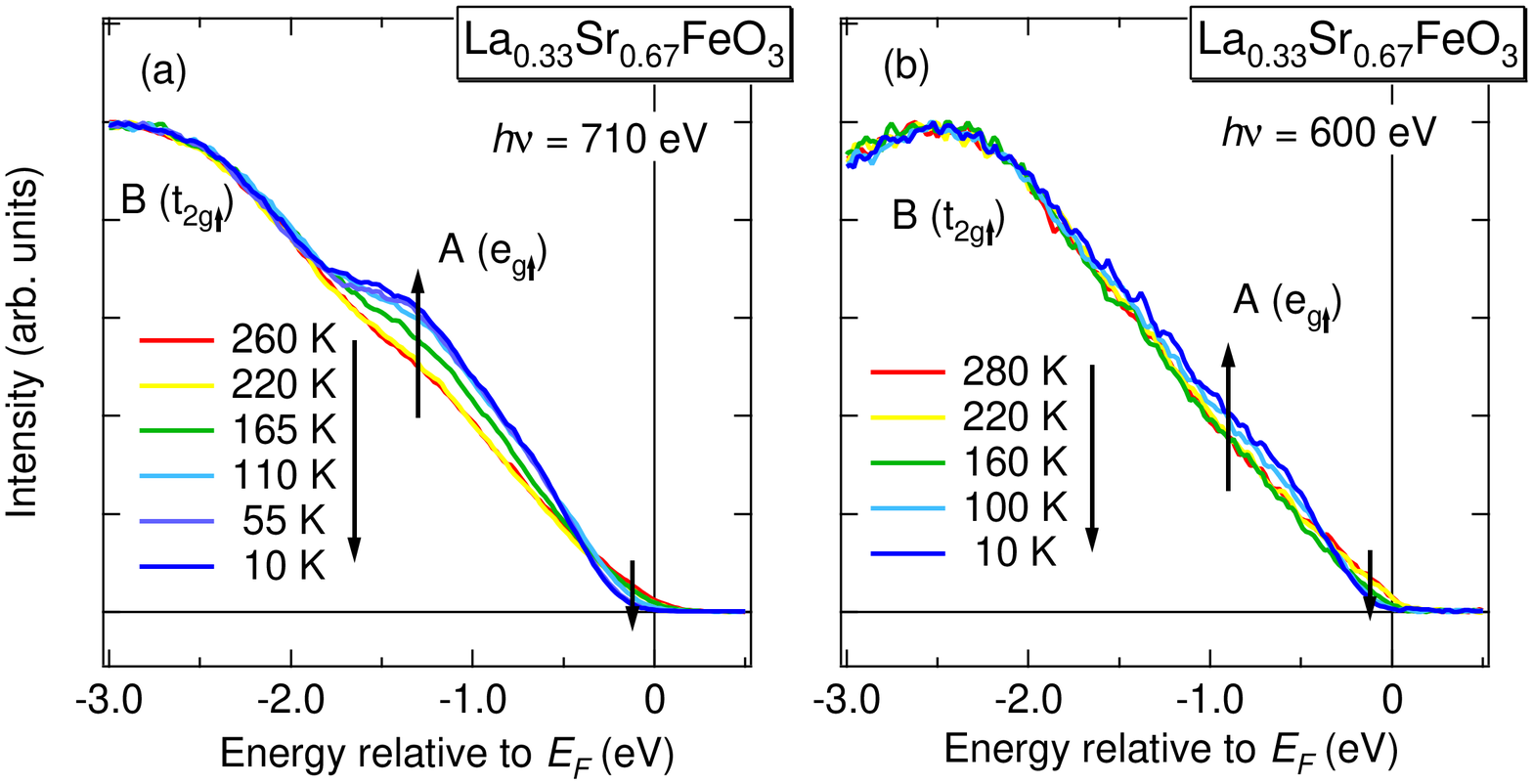}
\includegraphics[width=8.5cm]{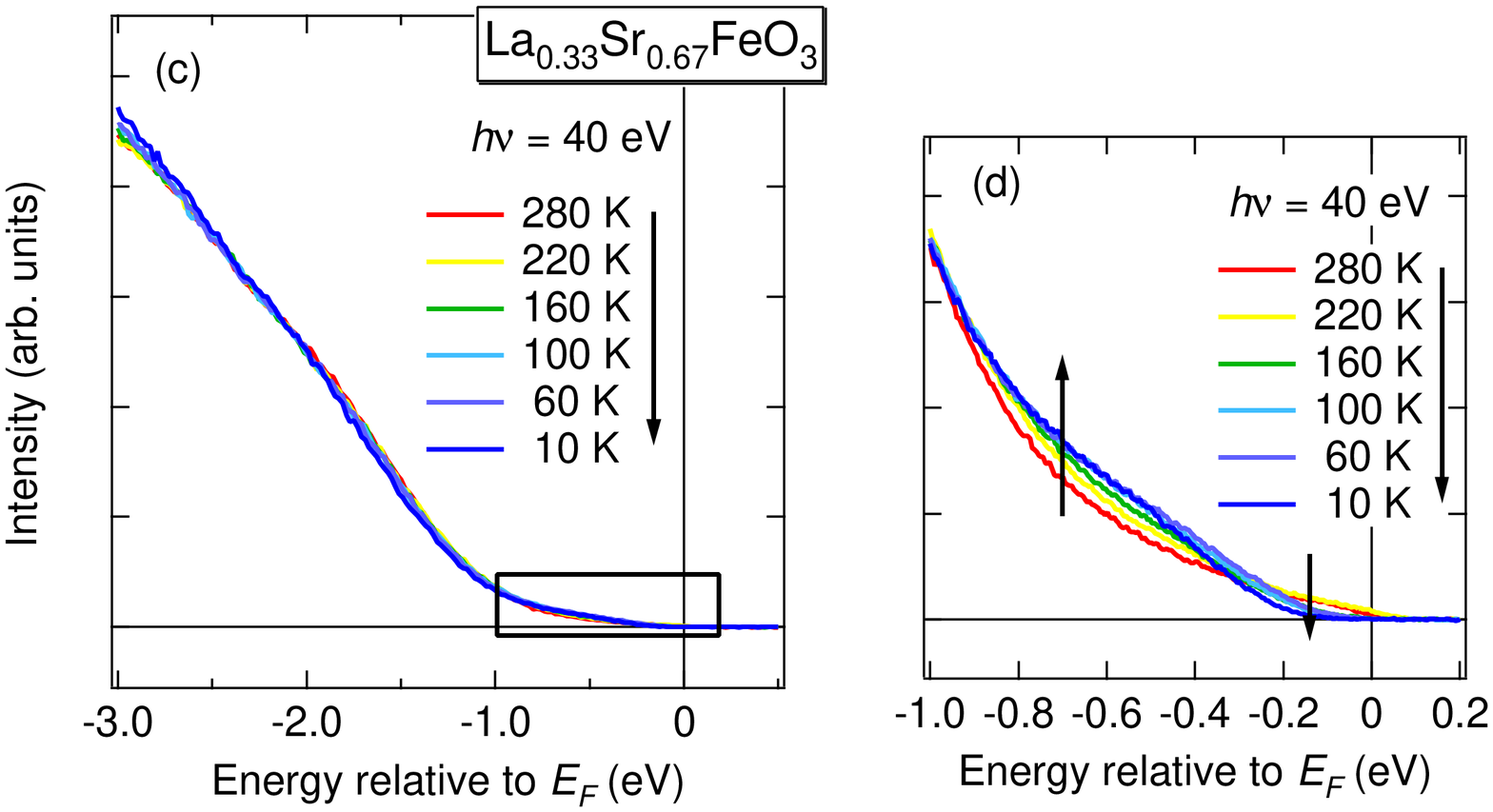}
\caption{(Color) Temperature dependence of the valence-band photoemission 
spectra of La$_{0.33}$Sr$_{0.67}$FeO$_3$. 
(a): $h\nu = 710$ eV (on Fe $2p\rightarrow 3d$ resonance), 
(b): $h\nu = 600$ eV, 
(c) and (d): $h\nu = 40$ eV. (d) shows an enlarged plot of (c) near $E_F$.} 
\label{0.67}
\end{center}
\end{figure}

Figures \ref{0.671} and \ref{0.67} show the temperature dependence of the valence-band 
photoemission spectra of La$_{0.33}$Sr$_{0.67}$FeO$_3$ taken at 
$h\nu = 710$ eV (on Fe $2p\rightarrow 3d$ resonance), 
600 eV and 40 eV. 
The spectra taken at $h\nu = 710$ eV are considered 
to represent the Fe $3d$ partial DOS (PDOS) with extra 
enhancement of the satellite around $-11$ eV, whereas 
those taken at $h\nu = 40$ eV are dominated by the O $2p$ PDOS and 
those taken at $h\nu = 600$ eV are 
mixtures of the Fe $3d$ and O $2p$ PDOS according to 
the corresponding atomic orbital photoionization cross-sections 
\cite{table}. 
In Fig.~\ref{0.671}, the spectra have been 
normalized to the area from 1 eV to $-14$ eV. 
One can observe four structures, namely, 
A ($e_{g\uparrow}$ band), 
B ($t_{2g\uparrow}$ band), 
C (Fe $3d-$O $2p$ bonding 
states), and the satellite structure. 
The intensity at $E_F$ is considerably low without a clear Fermi-level
cutoff even at room temperature. 
A similar observation was reported for Fe$_3$O$_4$ \cite{mag1,mag2,Fe3O4}, 
which also shows charge ordering below $\sim$ 120 K. 
The spectra change gradually with temperature. 
With decreasing temperature, spectral weight 
is transferred from $-(4-5)$ eV to structure A ($e_{g\uparrow}$ band) 
while the height of 
structure B ($t_{2g\uparrow}$ band) remains almost unchanged. 
In Fig.~\ref{0.67} (a) (b) and thereafter 
we have normalized 
the spectra to the height of structure B, and concentrate on 
the spectra within $\sim 3$ eV of $E_F$. 
The spectra for 710 eV and 600 eV clearly indicate a 
transfer of spectral weight from the vicinity of $E_F$ 
to structure A, that is, within the $e_{g\uparrow}$ band region 
with decreasing temperature. 
The energy range in which spectral weight transfer occurs is 
two orders of magnitude larger than 
the transition temperature $T_{CD}=190$ K ($\sim 22$ meV). 
Spectral weight transfer over such a wide energy range was also reported 
in other transition-metal oxides such as 
La$_{1-x}$Sr$_{x}$MnO$_3$ \cite{Sarma2}, 
La$_{1-x}$Ca$_{x}$MnO$_3$ \cite{Park1}, 
Nd$_{1-x}$Sr$_{x}$MnO$_3$ \cite{Suga1},
Pr$_{1-x}$Sr$_{x}$MnO$_3$ \cite{Chainani2} 
and Nd$_{1-x}$Sm$_x$NiO$_3$ \cite{Vobornik,Okazaki}. 
The 40 eV spectrum [Fig.~3 (c)] does not show appreciable changes 
on the same energy and intensity scales, 
but near $E_F$ a clear spectral weight transfer is 
observed [Fig.~3 (d)]. 
From those photon energy dependences, we conclude that 
the temperature-dependent part of O $2p$ states are distributed 
within $\sim 0.8$ eV of $E_F$ and 
the temperature-dependent part of 
Fe $3d$ states within $\sim 2$ eV of $E_F$.

\begin{figure}
\begin{center}
\includegraphics[width=9.5cm]{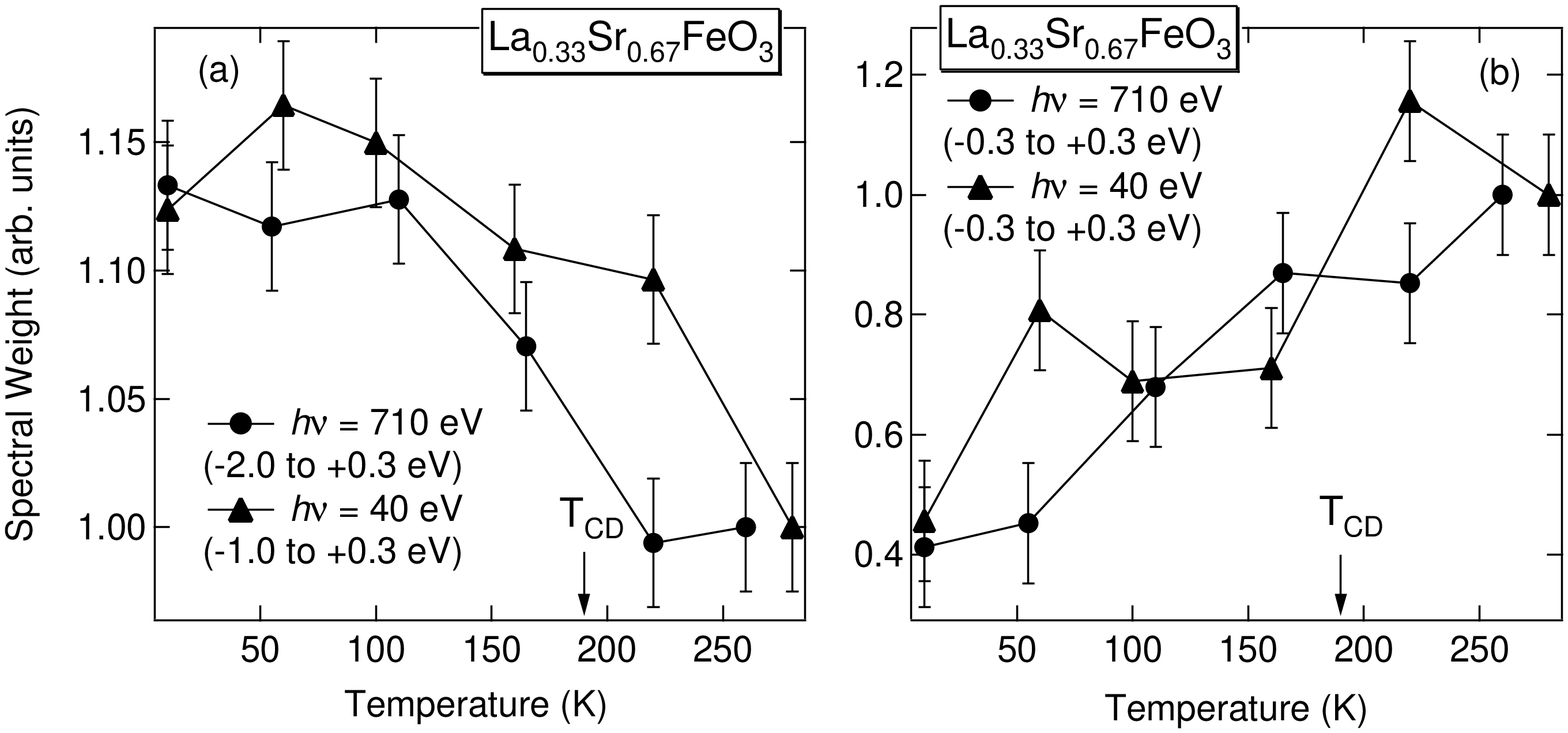}
\caption{Spectral weight in the $e_{g\uparrow}$ band region 
for La$_{0.33}$Sr$_{0.67}$FeO$_3$ as a function of temperature : 
(a) Spectral weight integrated from $-2.0$ eV to $+0.3$ eV 
for $h\nu = 710$ eV (representing the Fe $3d$ PDOS) and 
from $-1.0$ eV to $+0.3$ eV for $h\nu = 40$ eV (representing the O $2p$ PDOS). 
(b) Spectral weight near $E_F$ integrated from $-0.3$ eV to $+0.3$ eV 
for $h\nu = 710$ eV and $40$ eV} 
\label{weight}
\end{center}
\end{figure}

In order to characterize the spectral weight transfer more clearly, 
we have plotted 
the spectral weight of the 710 eV spectrum (Fe $3d$ PDOS) 
integrated from $-2.0$ eV to $+0.3$ eV and 
that of the 40 eV spectrum (O $2p$ PDOS) 
integrated from $-1.0$ eV to $+0.3$ eV 
as a function of temperature 
in Fig.~\ref{weight} (a). We have also plotted the spectral weight 
integrated from $-0.3$ eV to $+0.3$ eV for the same  photon energies 
in Fig.~\ref{weight} (b). A dramatic but gradual change of spectral weight 
is observed across the transition temperature in both plots. 
While the spectral weight within $\sim 2$ eV of $E_F$ 
increases with decreasing temperature, 
reflecting the spectral weight transfer from $-(4-5)$ eV 
to structure A, spectral weight in the vicinity of $E_F$ 
decreases with decreasing temperature, 
reflecting the spectral weight transfer from near $E_F$ 
to structure A.

\begin{figure}
\begin{center}
\includegraphics[width=9.5cm]{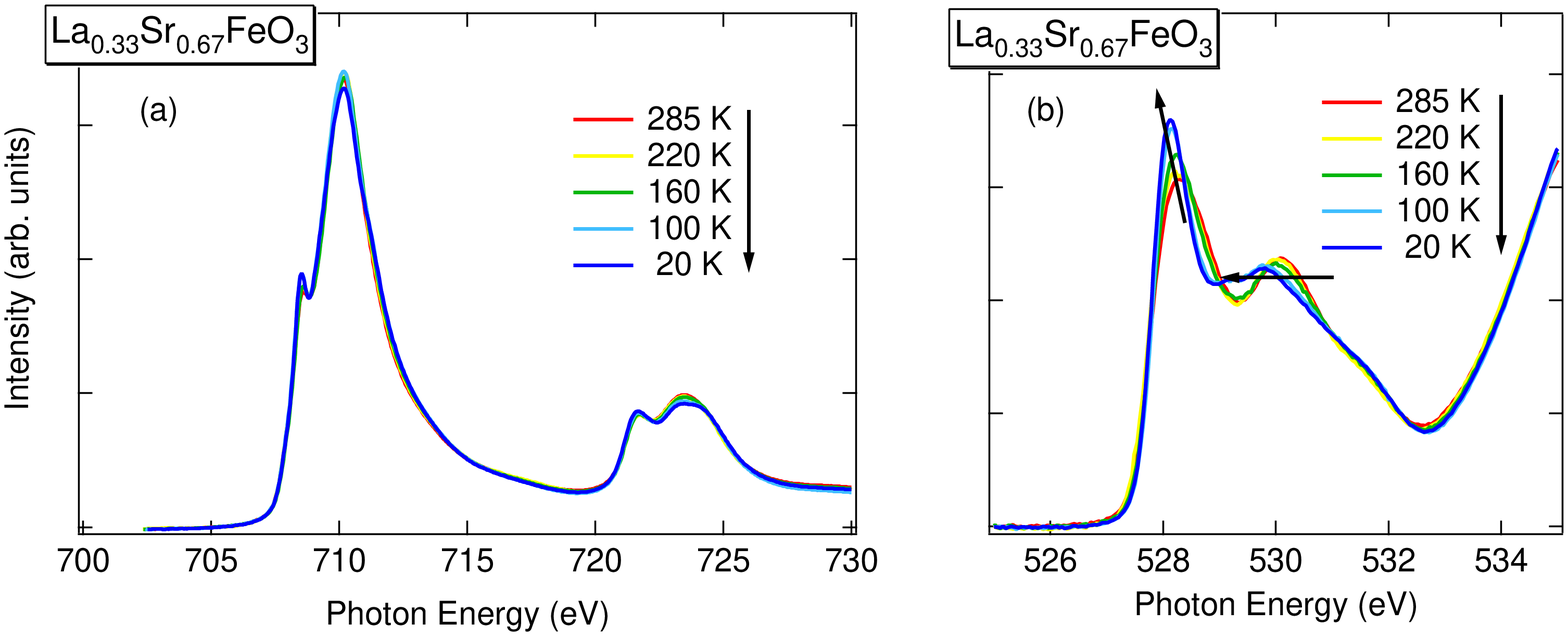}
\caption{(Color) Temperature dependence of XAS spectra of 
La$_{0.33}$Sr$_{0.67}$FeO$_3$. (a) Fe $2p$ (b) O $1s$.} 
\label{XAS1}
\end{center}
\end{figure}

Next, 
Fig.~\ref{XAS1} shows the temperature dependence of 
the Fe $2p$ and O $1s$ XAS spectra. 
One cannot observe any changes in the Fe $2p$ XAS spectra 
within experimental errors, whereas there is a significant 
change in the O $1s$ XAS spectra. 
Upon cooling, the peaks in the empty Fe $3d$ band 
region become sharp, 
and the first peak corresponding to the doping-induced $e_{g\uparrow}$ band 
becomes more intense. 
This spectral change is opposite to the change of 
the photoemission spectra in the vicinity of $E_F$ but is the same as 
the increase of structure A. 
At the same time, spectral weight on the higher energy side of 
the first peak decreases upon cooling, resulting in the overall shift of
the first peak toward lower energies. In addition, the structure at
$\sim$ 530 eV changes its lineshape and is shifted to lower energies.

\begin{figure}
\begin{center}
\includegraphics[width=8cm]{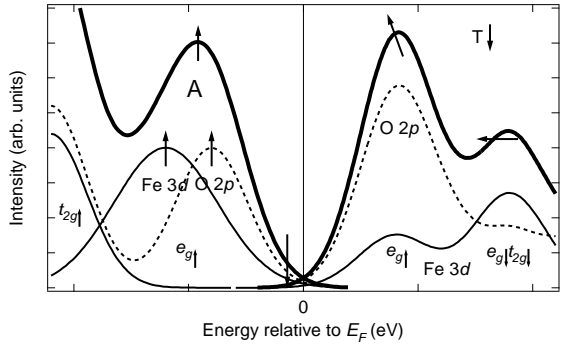}
\caption{Schematic picture of the spectral changes with decreasing
 temperature in La$_{0.33}$Sr$_{0.67}$FeO$_3$.} 
\label{band067}
\end{center}
\end{figure}

To summarize those spectral changes, we show 
a schematic picture in Fig.~\ref{band067}. 
Below $E_F$, upon cooling 
spectral weight is transferred 
from the near-$E_F$ states as well as from the states
at $-(4-5)$ eV of Fe $3d$ character 
to the states of Fe $3d$ character and O $2p$ character at 
$\sim -2$ eV and $\sim -1$ eV, respectively 
(composing structure A). 
Correspondingly, there is an increase in the O $2p$ DOS at 
$\sim +1$ eV above $E_F$ upon cooling. We believe that the DOS in the
vicinity of $E_F$ decreases above $E_F$, too, 
since a gap should open at $E_F$, 
but the change associated with the gap opening may be too small 
to be observed with the resolution of O $1s$ XAS. 

The gradual temperature dependence indicates that CD does not occur
abruptly at $T_{CD}$ but occurs gradually from well above $T_{CD}$, 
namely, that a local CD occurs well above $T_{CD}$ and 
continues to develop below $T_{CD}$, too. 
It would be interesting to investigate other Fe oxides which undergoes
charge ordering such as CaFeO$_3$ and Fe$_3$O$_4$ to see whether such a
temperature-dependent 
spectral weight transfer is a common phenomenon or not. 
\subsection{La$_{1-x}$Sr$_x$FeO$_3$ with $x=0,\ 0.2,\ 0.4$}
We have also measured LSFO thin films with $x=0$, 0.2, and 0.4 
by photoemission and O $1s$ XAS 
to see whether the temperature dependence seen for $x=0.67$ 
is unique to this composition or not. 
For $x=0$, 0.2, and 0.4, 
the system is insulating at all temperatures 
without any abrupt change
in the resistivity as shown in Fig.~\ref{pt} 
and exhibits only antiferromagnetic transitions 
above room temperature (at about 740 K, 520 K, 
and 320 K, respectively). 

\begin{figure}
\begin{center}
\includegraphics[width=9cm]{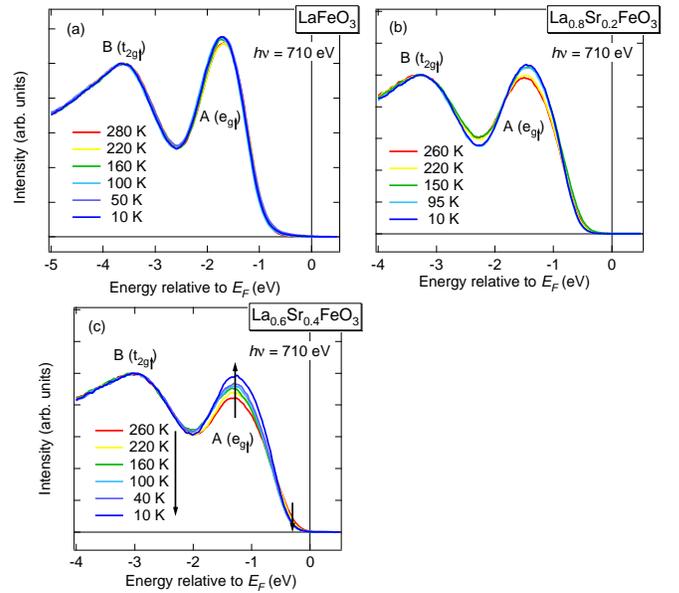}
\caption{(Color) Temperature dependence of the valence-band photoemission spectra of
 La$_{1-x}$Sr$_x$FeO$_3$ taken at $h\nu =$ 710 eV, i.e. Fe $2p\rightarrow 
3d$ resonance. (a) $x=0$ (b) $x=0.2$ (c) $x=0.4$.} 
\label{708eV}
\end{center}
\end{figure}

In Fig.~\ref{708eV}, we show the temperature dependence of the valence-band 
photoemission spectra of LSFO with $x=0$, 0.2, and 0.4 taken at 
$h\nu = 710$ eV, i.e., on Fe $2p\rightarrow 3d$ resonance. 
The intensity of peak A increases with decreasing $x$ 
due to the increase in the $e_{g\uparrow}$ band filling, as reported in the
previous paper \cite{Wadati}. 
There was only very weak temperature dependence for $x=0$, as expected. 
As for $x=0.2$ and 0.4, however, the spectra change gradually 
with temperature as in the case of $x=0.67$, that is, 
the intensity of peak A increases and that in the vicinity of $E_F$ 
decreases with decreasing temperature. 
Surprisingly, the magnitude of the changes for $x=0.4$ 
is comparable to that of $x=0.67$. 

\begin{figure}
\begin{center}
\includegraphics[width=9cm]{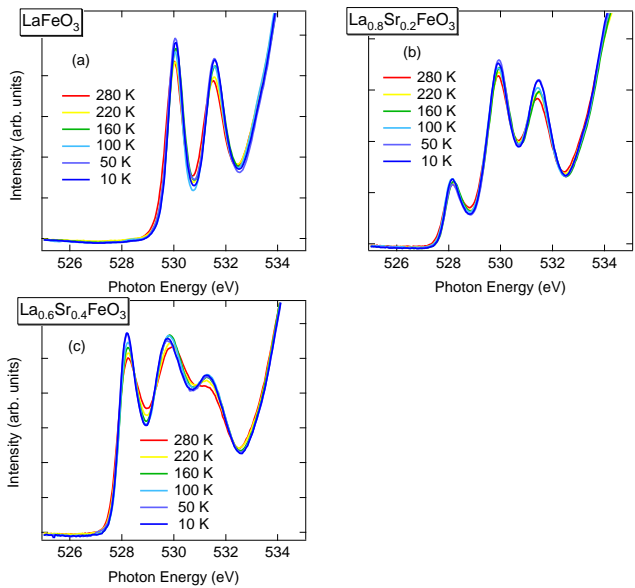}
\caption{(Color) Temperature dependence of the O $1s$ XAS spectra 
of La$_{1-x}$Sr$_x$FeO$_3$. (a) $x=0$ (b) $x=0.2$ (c) $x=0.4$.} 
\label{XAS}
\end{center}
\end{figure}

We also measured the temperature dependence of the O $1s$ XAS spectra, 
as shown in Fig.~\ref{XAS}. The spectra for all $x$'s show 
appreciable temperature dependence. 
Particularly for $x=0.4$, 
the change is as dramatic as in the case
of $x=0.67$. 
From Figs.~\ref{708eV} and \ref{XAS}, we conclude 
that the temperature dependence 
becomes stronger with increasing $x$ until $x=0.67$ 
but that $x=0.4$ shows as strong temperature 
dependence as $x=0.67$. 

Although there is no long-range charge ordering 
in the $x=0.4$ sample, 
the temperature dependence comparable to $x=0.67$ 
suggests that a local charge 
disproportionation occurs at low temperatures for $x=0.4$, too. 
The present results together with 
the previous report on the temperature-dependent 
photoemission study of $x=0.55$ and 0.80 \cite{Matsuno} suggest 
that the local charge disproportionation occurs 
over a wide composition range around $x=0.67$. 
If all Fe$^{4+}$ (predominantly of the $d^5\underline{L}$ 
configuration) 
ions in the $\mbox{Fe}^{4+}-\mbox{Fe}^{3+}$ mixed valence state 
disproportionate into Fe$^{3+}$ ($d^5$) and 
Fe$^{5+}$ ($d^5\underline{L}^2$), the degree of the temperature dependence 
would be proportional to the concentration of the Fe$^{4+}$ ions and 
hence to $x$, which may be consistent with the result 
that the spectral change of $x=0.4$ is comparable to 
that of $x=0.67$. 
The M\"{o}ssbauer spectra of LSFO with $x=0.5$ \cite{Yoon}, which show 
valence changes similar to the case of $x=0.67$, 
support our scenario of local charge disproportionation. 
For $x>0.67$, the local charge-disproportionated 
state may become unstable because this system approaches SrFeO$_3$, 
where the metallic state is stable. 

One striking feature of the LSFO phase diagram 
is that the insulating phase is unusually 
wide ($0<x<0.9$ at low temperatures, 
and $0<x<0.5$ at room temperature) \cite{Matsuno}. 
The observation that 
there is a gap at $E_F$ for all values of $x$ in the
photoemission and XAS spectra \cite{Wadati} 
is naturally related to the wide insulating phase of this material. 
From the present work, we propose that 
the wide insulating region with the gap or pseudogap 
are due to the local 
charge disproportionation. 
Search for, e.g., diffuse scattering corresponding to the local 
charge disproportionation in samples with $x\ne 0.67$ 
is desired to test our proposal. 
\section{Conclusion}
We have measured the temperature dependence of the photoemission and XAS 
spectra of La$_{0.33}$Sr$_{0.67}$FeO$_3$ to study the spectral 
change across the transition temperature of charge disproportionation. 
We observed gradual changes of the spectra 
with temperature. 
Above $T_{CD}$ the intensity at the Fermi level ($E_F$) 
becomes relatively high compared to that below $T_{CD}$ 
but shows no clear Fermi-level
cutoff and still remains much lower than that in 
conventional metals. 
Below $E_F$, spectral weight transfer was observed. 
Spectral weight is transferred 
from the near-$E_F$ states of mixed 
Fe $3d-$O $2p$ character 
to the states of both Fe $3d$ and 
O $2p$ character well below $E_F$ primarily within the $e_{g\uparrow}$ 
band. 
Strong temperature dependence was also observed above $E_F$ in the 
O $1s$ XAS spectra. 
We also measured the temperature dependence of the photoemission and XAS 
spectra of La$_{1-x}$Sr$_x$FeO$_3$ with $x=0$, 0.2, 0.4. 
For $x=0.4$, temperature dependence comparable to $x=0.67$ 
was observed. 
Those observations suggest that a local charge disproportionation 
occurs not only in the $x=0.67$ sample below 190 K but over a wider
temperature and composition range. 
We propose that 
this local charge disproportionation is 
the origin of the wide insulating region of the LSFO phase diagram.  

\section*{Acknowledgment}
The authors would like to thank D. Kobayashi for sample preparation 
and T. Mizokawa for informative discussions. 
We are also grateful to K. Ono and A. Yagishita for their support 
at KEK-PF. 
This work was supported by a Grant-in-Aid for Scientific Research 
(A16204024) from the Ministry of Education, Culture, Sports, 
Science and Technology, Japan.
This work was done under the approval of the Photon Factory 
Program Advisory Committee (Proposal No. 2003G149).
\bibliography{LSFOtex}
\end{document}